\documentclass[apj]{emulateapj}

\usepackage{amsmath}
\usepackage{verbatim}
\usepackage{natbib}
\usepackage{enumerate}

\newcommand{\der}[2]{\frac{d #1}{d #2}}

\newcommand{\pder}[2]{\left(\frac{\partial #1}{\partial #2}\right)}

\begin{document}
\title{Main Sequence Evolution with Layered Semiconvection}
\author{Kevin Moore \altaffilmark{1,2,3}}
\author{Pascale Garaud \altaffilmark{1}}
\altaffiltext{1}{Department of Applied Mathematics and Statistics, University of California, Santa Cruz, CA}
\altaffiltext{2}{Department of Astronomy \& Astrophysics, University of California, Santa Cruz, CA}
\altaffiltext{3}{TASC, University of California, Santa Cruz, CA}

\begin{abstract}
Semiconvection - mixing that occurs in regions that are stable when considering compositional gradients, but unstable when ignoring them - is shown to have the greatest potential impact on main sequence stars with masses in the range $1.2\ M_\odot - 1.7\ M_\odot$. We present the first stellar evolution calculations using a prescription for semiconvection derived from extrapolation of direct numerical simulations of double-diffusive mixing down to stellar parameters. The dominant mode of semiconvection in stars is layered semiconvection, where the layer height is an adjustable parameter analogous to the mixing length in convection. The rate of mixing across the semiconvective region is sensitively dependent on the layer height. We find that there is a critical layer height that separates weak semiconvective mixing (where evolution is well-approximated by using the Ledoux criterion) from strong semiconvective mixing (where evolution is well-approximated by using the Schwarzschild criterion). This critical layer height is much smaller than the minimum layer height expected from simulations so we predict that for realistic layer heights, the evolution is nearly the same as a model ran with the Schwarzschild criterion. We also investigate the effects of compositional gradient smoothing, finding that it causes convective cores to artificially shrink in the absence of additional mixing beyond the convective boundary. Layered semiconvection with realistic layer heights provides enough such mixing that stars will still evolve as if the Schwarzschild criterion is employed. Finally, we discuss the potential of detecting such semiconvection and its implication on convective core sizes in solar-like oscillators.
\end{abstract}


\maketitle

\section{Introduction}
\label{sec:intro}
One of the outstanding challenges in constructing accurate stellar models is understanding macroscopic mixing driven by fluid instabilities. 
Direct numerical simulations of fluid instabilities, however, cannot be included in stellar evolution calculations over an appreciable fraction of a star's life, due to the large separation of length and time scales between small-scale fluid motion and global stellar behavior. Instead, stellar evolution codes rely on one-dimensional prescriptions for the transport of energy, composition, and angular momentum associated with instabilities such as convection, overshoot, shear, etc.
Mixing length theory (MLT) \citep{Bohm-Vitense58} is the standard way of modeling convection in stars and planets, and is usually applied in two steps. First, one determines which regions are convectively stable and unstable, then one adds a model for the convective flux to the total heat flux as well as a model for turbulent mixing in the evolution equation for each chemical species. The first of these two steps is typically done in a local manner through a comparison of structural gradients within the object \citep[see also][]{KW}. Ignoring composition gradients, the presence or absence of convection is established by the \textit{Schwarzschild criterion}, which states that a region is stable when
\begin{equation}
\frac{3\kappa P l}{16\pi a c G T^4 m} \equiv \nabla_{\rm rad} < \nabla_{\rm ad} \equiv \pder{\ln T}{\ln P}_{\rm ad},
\end{equation}
where $\nabla_{\rm rad}$ is the radiative temperature gradient (the temperature gradient required if all the energy is transported via photon diffusion), $\nabla_{\rm ad}$ is the adiabatic temperature gradient, computed directly from the equation of state, $\kappa$ is the opacity, $P$ is the pressure, $l$ is the luminosity, $a$ is the radiative constant, $c$ the speed of light, $G$ the gravitational constant, $T$ the temperature, and $m$ the local mass coordinate. To account for composition gradients, one must use the \textit{Ledoux criterion} instead, whereby stability occurs when
\begin{equation}
\nabla_{\rm rad} < \nabla_{\rm ad} + \frac{\phi}{\delta}\nabla_\mu \equiv \nabla_L,
\end{equation}
where $\phi$ and $\delta$ are thermodynamic derivatives of the equation of state,
\begin{equation}
\phi = \pder{\ln \rho}{\ln \mu}_{P,T}, \hspace{1cm} \delta = -\pder{\ln \rho}{\ln T}_{P,\mu}
\end{equation}
and 
\begin{equation}
\nabla_{\mu} = \der{\ln \mu}{\ln P}
\end{equation}
is the local non-dimensional composition gradient, with $\mu$ being the mean molecular weight of the material. We will refer to stellar models computed using the Schwarzschild criterion as the Schwarzschild case or Schwarzschild models, and models computed using the Ledoux criterion will be similarly named.

If the composition gradient is stabilizing $(\nabla_\mu > 0,\ {\rm so}\ \nabla_{\rm ad} < \nabla_L)$ then there can be regions where $\nabla_{\rm ad} < \nabla_{\rm rad} < \nabla_L$ that are stable to convection under the Ledoux criterion, but unstable under the Schwarzschild criterion. It is now known \citep{Kato66} that such regions can be subject to a different kind of instability called oscillatory double-diffusive convection (ODDC) and are often referred to as semiconvective \citep{Schwarzschild58}. Being diffusive in nature, semiconvective mixing is thought to be much weaker than convective mixing, and is typically modeled with a separate prescription \citep{Langer85, Castellani85}, the strength of which is controlled by additional parameters analogous to the mixing length in convection. 

Semiconvection was historically used in modeling high-mass stars $\ge 10\ M_\odot$ \citep{Schwarzschild58, Stothers70, Stothers75, Langer85}. Semiconvective mixing has been investigated for lower-mass stars on the main sequence \citep{Faulkner73, Gabriel77, Crowe82, Silva-Aguirre10a}, but not using prescriptions motivated by numerical simulations. Semiconvective mixing in main sequence stars is typically encountered during convective core burning, just outside of the region that is Ledoux-unstable to convection. It is caused by the buildup of composition gradients induced by the long tail of low-temperature proton-proton chain burning that extends outside of the convective core for main sequence stars in the mass range $1.2\ M_\odot - 1.7\ M_\odot$ (see section \ref{subsec:where_semi}). Generally speaking, they occur whenever there is a stabilizing composition gradient adjacent to a convection zone, such as near H-burning shells in evolved stars as well as in later stages of massive star evolution.



In this paper we implement the new prescription for semiconvective mixing proposed by \citet{Wood13} into the open-source stellar evolution code MESA\footnote{Version 6794} \citep{Paxton11, Paxton13}. We review the general properties of the \citeauthor{Wood13} model in section \ref{sec:double-diffusive}. We argue that ODDC most likely takes the form of layered semiconvection for parameters representative of stellar interiors \citep[see also][]{Garaud14}, with the layer height remaining a free parameter. Section \ref{sec:implementation} details how we implement this mixing prescription in MESA. It also discusses the effects of numerical smoothing on convective core evolution. Section \ref{sec:applications} presents applications of this theory to the growth of convective cores in main sequence stars and identifies the stellar mass range which is most sensitive to the effects of semiconvection. We show the effects of varying layer heights on convective core growth, and derive a critical layer height for semiconvective regions to persist throughout the main sequence. We summarize our conclusions in section \ref{sec:conclusions} and briefly outline future directions of investigation into semiconvective mixing.

\section{Oscillatory double-diffusive convection}
\label{sec:double-diffusive}
As discussed in section \ref{sec:intro}, oscillatory double diffusive convection is a mild form of convection that occurs in the presence of stabilizing chemical gradients and destabilizing thermal gradients.
The onset and strength of ODDC is controlled by several dimensionless parameters. The first is the Prandtl number,
\begin{equation}
{\rm Pr} = \frac{\nu}{\kappa_T},
\end{equation}
which is the ratio of the microscopic kinematic viscosity $\nu$ to the thermal diffusivity $\kappa_T$ (both with units of cm$^2$/s in cgs units). Low Prandtl numbers ($\approx 10^{-8} - 10^{-3}$) are typical in stars and gas giant planets, while telluric planet interiors may have ${\rm Pr} > 1$ \citep{Soderlund13}.
The second parameter is the diffusivity ratio,
\begin{equation}
\tau = \frac{\kappa_\mu}{\kappa_T},
\end{equation}
of the microscopic compositional diffusivity $\kappa_\mu$, to the thermal diffusivity. ODDC only occurs when 
$\tau < 1$, which is the standard situation in stars since heat is transported via photon diffusion as well as collisions between nuclei, while chemical species only diffuse through collisional processes. Finally, the third parameter is the density ratio
\begin{equation}
R_0 = \frac{\delta (\nabla - \nabla_{\rm ad})}{\phi \nabla_\mu},
\end{equation}
where $\delta$ and $\phi$ are the equation of state derivatives defined in the previous section (with $\delta = \phi = 1$ for an ideal gas).
As discussed by \citet{Baines69}, semiconvection occurs when
\begin{equation}
R_{\rm crit} < R_0 < 1,
\label{eq:semiconvection_criterion}
\end{equation}
where
\begin{equation}
R_{\rm crit} = \frac{{\rm Pr} + \tau}{{\rm Pr} + 1}.
\end{equation}
Equation (\ref{eq:semiconvection_criterion}) is a local criterion, analogous to the Schwarzschild and Ledoux criteria used to determine convective instability. It can be rewritten in terms of $\nabla$ as
\begin{equation}
\nabla_{\rm ad} + R_{\rm crit}\frac{\phi}{\delta}\nabla_{\mu} < \nabla < \nabla_{\rm ad} + \frac{\phi}{\delta}\nabla_{\mu}.
\label{eq:semi_criterion}
\end{equation}
For ${\rm Pr}, \tau \ll 1$ we have $R_{\rm crit} \ll 1$, so the semiconvective criterion here is nearly what is typically employed in stars, $\nabla_{\rm ad} < \nabla < \nabla_{\rm ad} + \frac{\phi}{\delta}\nabla_{\mu}$ \citep{Langer85}. Regions with $R_0 > 1$ ($\nabla > \nabla_{\rm ad} + \frac{\phi}{\delta}\nabla_{\mu}$) are unstable to convection, while regions with $R_0 < R_{\rm crit}$ ($\nabla < \nabla_{\rm ad} + R_{\rm crit}\frac{\phi}{\delta}\nabla_{\mu}$) are stable and therefore radiative.

The region of parameter space unstable to ODDC is itself divided into two distinct domains, one in which the semiconvection later transitions into a layered state, and one in which it does not. The former has been determined empirically through simulations by \citet{Mirouh12}, to encompass virtually all of the ODDC region for stellar parameter regimes of ${\rm Pr}, \tau \ll 1$. In what follows, we therefore ignore the possibility of non-layered semiconvection. Layered semiconvection takes the form of fully convective layers of height $H_L$, separated by thin, stably stratified interfaces. For stellar parameters, the interfaces are very mobile, and tend to merge rapidly with one another. As a result, the ultimate layer height $H_L$ in semiconvection is difficult to determine from numerical simulations and will be treated as a free parameter of the model, much like the mixing length in standard MLT.

The total heat flux at a given location in a star can be expressed as the sum of the radiative and semiconvective heat fluxes,
\begin{equation}
F_{\rm tot} = F_{\rm rad} + F_{\rm semi},
\label{eq:total_flux}
\end{equation}
where the radiative heat flux is given by
\begin{equation}
F_{\rm rad} = -k_{\rm rad} \der{T}{r},
\end{equation}
and the semiconvective heat flux is given by
\begin{equation}
F_{\rm semi} = -k_{\rm rad} ({\rm Nu}_T - 1)\left(\der{T}{r} - \left.\der{T}{r}\right|_{\rm ad}\right),
\label{eq:fsemi_orig}
\end{equation}
where $k_{\rm rad} = \rho c_P \kappa_T$ is the microscopic thermal conductivity and ${\rm Nu}_T$ is the thermal Nusselt number (see below).
Similarly, the turbulent compositional flux is written,
\begin{equation}
F_{\mu} = -\kappa_\mu ({\rm Nu}_\mu - 1) \der{\mu}{r},
\end{equation}
where ${\rm Nu}_\mu$ is the compositional Nusselt number. 
\citet{Wood13} ran a systematic set of direct numerical simulations of layered semiconvection to measure the Nusselt numbers ${\rm Nu}_T$ and ${\rm Nu}_\mu$ as functions of the model parameters $R_0$, Pr, $\tau$, and $H_L$. They found that both Nusselt numbers vary only weakly with $R_0$, and depend on $H_L$ only through the Rayleigh number, Ra, given by
\begin{equation}
{\rm Ra} = \left| \frac{\delta}{\rho} \der{P}{r} \left(\der{\ln T}{r} - \der{\ln T_{\rm ad}}{r}\right) \frac{H_L^4}{\kappa_T \nu} \right|.
\end{equation}
Empirical fits to the data suggest that 
\begin{equation}
{\rm Nu}_T - 1 = 0.1\ {\rm Pr}^{1/3} {\rm Ra}^{1/3},
\label{eq:nu_t}
\end{equation}
and
\begin{equation}
{\rm Nu}_\mu - 1 = 0.03\ \tau^{-1} {\rm Pr}^{1/4} {\rm Ra}^{0.37}.
\label{eq:nu_mu}
\end{equation}
The next section details how this mixing prescription is actually implemented in MESA.

\section{Implementation in MESA}
\label{sec:implementation}
\subsection{Semiconvective mixing prescription}
MESA is a one-dimensional Lagrangian code that solves the equations of stellar structure in a fully coupled (unsplit) method, although options to split the mixing equations from the burning and structure equations exist. The two equations that are directly affected by the addition of extra mixing are the equations for thermal and compositional transport. The first is discretized as
\begin{equation}
T_{k-1} - T_k = \overline{dm}_k\left[ \nabla_k \left(\der{P}{m}\right)\frac{\overline{T}_k}{\overline{P}_k} \right],
\end{equation}
where $T_k$ is the temperature at the center of cell $k$, $\overline{T}_k$ is the mass-interpolated temperature at the outer face of cell $k$ (similar for pressures $P_k$ and $\overline{P}_k$), $\overline{dm}_k$ is the mean mass of cells $k$ and $k-1$, and $\nabla_k$ is the temperature gradient at the face of cell $k$ (cell indices in MESA increase inward towards the core). While not explicit in this equation, the semiconvective prescription directly affects the the calculation of the local temperature gradient, $\nabla$ (see section \ref{sec:nabla} below). The equation of compositional transport on the other hand, is
\begin{equation}
X_{i,k-1}(t + \delta t) - X_{i,k}(t) = \der{X_{i,k}}{t}\delta t + \left( F_{i,k+1} - F_{i,k} \right)\frac{\delta t}{\overline{dm}_k},
\end{equation}
where $\delta t$ is the time step and
\begin{equation}
F_{i,k} = \left( X_{i,k} - X_{i, k-1} \right)\frac{D_k}{\overline{dm}_k}
\end{equation}
is the flux of the $i^{\rm th}$ species through the outer face of the $k^{\rm th}$ cell, $D_k$ is the compositional diffusion coefficient, and $X_{i,k}$ is the mass fraction of the $i^{\rm th}$ species in the $k^{\rm th}$ cell, see section 6.2 in \citet{Paxton11} for a more detailed discussion. In this case, semiconvective mixing directly affects the calculation of $D_k$. 

The mixing type of a given cell is determined entirely locally, with regions labeled as semiconvective when 
\begin{equation}
\nabla_{\rm ad} < \nabla_{\rm rad} < \nabla_{\rm ad} + \frac{\phi}{\delta}\nabla_{\mu}.
\end{equation}

\subsubsection{Temperature gradient calculation}
\label{sec:nabla}
As in standard MLT, we use the relationship between the heat flux and $\nabla$ to solve for the latter.
From equation (\ref{eq:fsemi_orig}), we rewrite the semiconvective flux as,
\begin{equation}
F_{\rm semi} = 0.1\ \frac{T\rho^2 c_P g \kappa_T^{1/3}}{P} \left| \frac{\delta \rho g^2}{P} \left(\nabla - \nabla_{\rm ad}\right) H_L^4 \right|^{1/3} \left(\nabla - \nabla_{\rm ad}\right),
\label{eq:f_semi}
\end{equation}
where $c_P$ is the specific heat at constant pressure, and $g=Gm/r^2$ is the local gravitational acceleration. Dividing equation (\ref{eq:total_flux}) by $k_{\rm rad} T\rho g/P$ yields,
\begin{equation}
\nabla_{\rm rad} = \nabla + \nabla_{\rm semi},
\label{eq:nabla_eqn}
\end{equation}
where
\begin{equation}
\nabla_{\rm semi} \equiv -\frac{F_{\rm semi}}{k_{\rm rad} T} \left(\der{\ln P}{r}\right)^{-1} = \frac{P}{T\rho g  k_{\rm rad}} F_{\rm semi}.
\label{eq:nabla_semi}
\end{equation}
Equation (\ref{eq:nabla_eqn}) is a fourth-order polynomial equation in $\nabla$ so could in principle be solved analytically for the latter. However, the expression of the solution is too complicated to be of practical value. Instead, we solve it numerically as part of the overall Newton solve to advance the star in time. The Jacobian for this Newton solve also requires partial derivatives of $\nabla$ with respect to local stellar variables, such as $P$, $T$, $m(r)$, etc. We calculate these by differentiating equation (\ref{eq:nabla_eqn}) implicitly. 

\subsubsection{Compositional mixing coefficient calculation}
\label{sec:D}
The compositional mixing rate is locally determined by the semiconvective diffusion coefficient, $D_{\rm semi}$. Its value in layered semiconvection is given by
\begin{align}
D_{\rm semi} &= \kappa_\mu({\rm Nu}_\mu - 1) \nonumber \\
&= 0.03 \frac{\kappa_T^{0.38}}{\nu^{0.12}} \left| \frac{\delta \rho g^2}{P} \left(\nabla - \nabla_{\rm ad}\right) H_L^4 \right|^{0.37},
\label{eq:d_layered}
\end{align}
and is calculated once $\nabla$ is known. As with $\nabla$, partial derivatives of $D_{\rm semi}$ are necessary for the Jacobian, which are computed by differentiating equation (\ref{eq:d_layered}).

\subsection{Effects of composition gradient smoothing on convective boundaries}
\label{subsec:smoothing}
When using the Ledoux criterion without additional mixing schemes such as semiconvection or overshoot, the resulting composition gradients are often not smooth. By default, MESA smooths $\nabla_\mu$ using a Gaussian smoothing formula with a 7-cell window. This kind of smoothing can have significant, but artificial effects on stellar evolution. Indeed, in core-convective stars, there is usually a sharp jump in composition between the outermost convective cell and the cell immediately above it. This produces a spike in $\nabla_\mu$ which grows in time, and is the main reason why 
in the absence of smoothing or any kind of mixing beyond the convective boundary, convective cores do not grow as large in the Ledoux case as compared to the Schwarzschild case. If this spike is smoothed before the MLT module computes the convective boundaries, then the region of high-$\nabla_\mu$ is artificially spread inward into cells that used to be convective. This then causes the convective boundary to move inward relative to that of a run which performs the same steps without compositional gradient smoothing. 

\begin{figure}
	\plotone{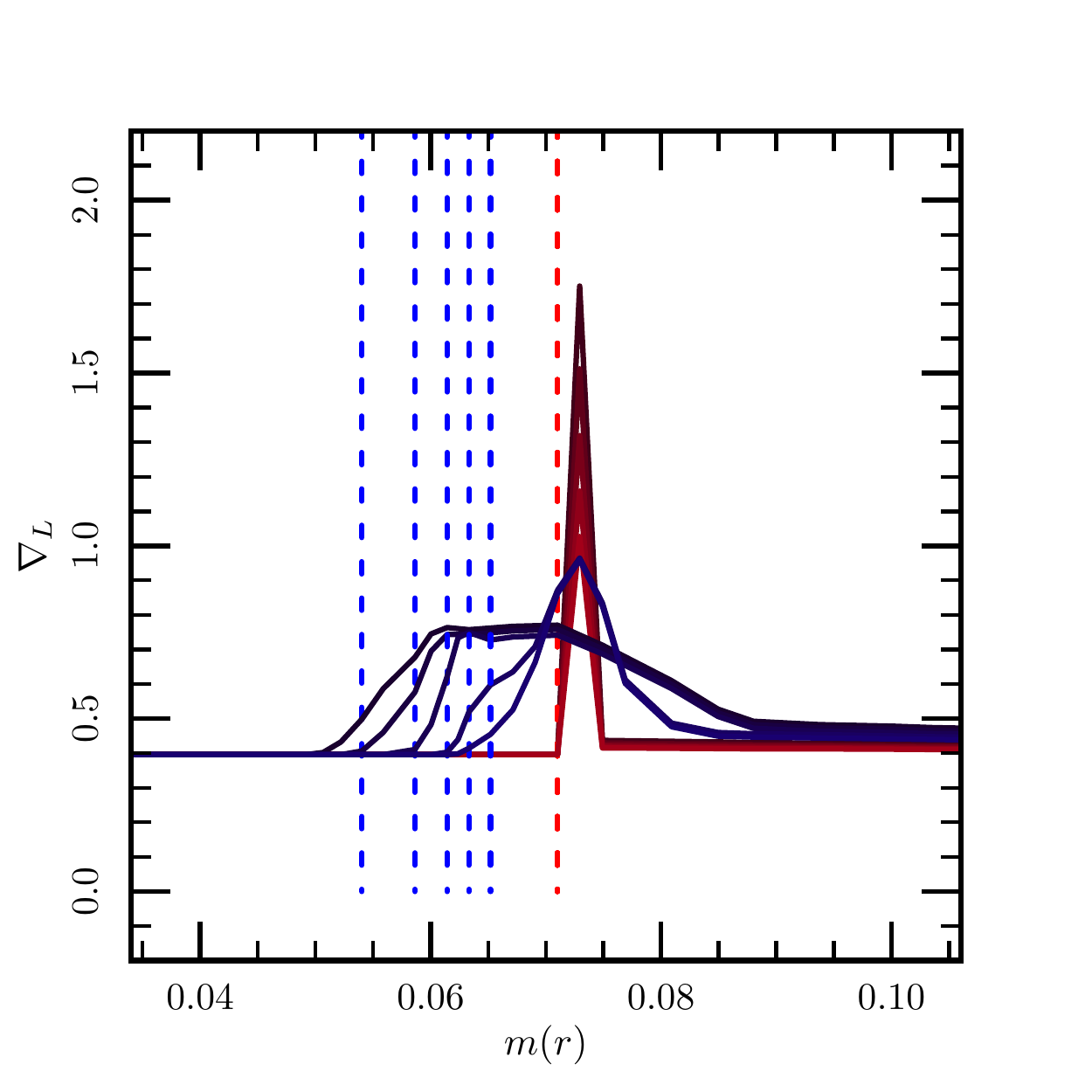}
	\caption{Evolution of $\nabla_L$ profiles of a $1.5\ M_\odot$ star when smoothing is suddenly turned on during evolution on the main sequence. Profiles of $\nabla_L$ as a function of mass coordinate are shown for five consecutive models where $\nabla_\mu$-smoothing is turned off (red), as well as five subsequent consecutive models with $\nabla_\mu$-smoothing enabled (blue). The extent of the convective core is shown by the dashed lines and is constant in mass coordinate until $\nabla_\mu$-smoothing is turned on, after which the outer mass coordinate of the convective core moves inward as the compositional gradient pushes into the formerly convective region.}
	\label{fig:smoothtest}
\end{figure}

\begin{figure}
	\plotone{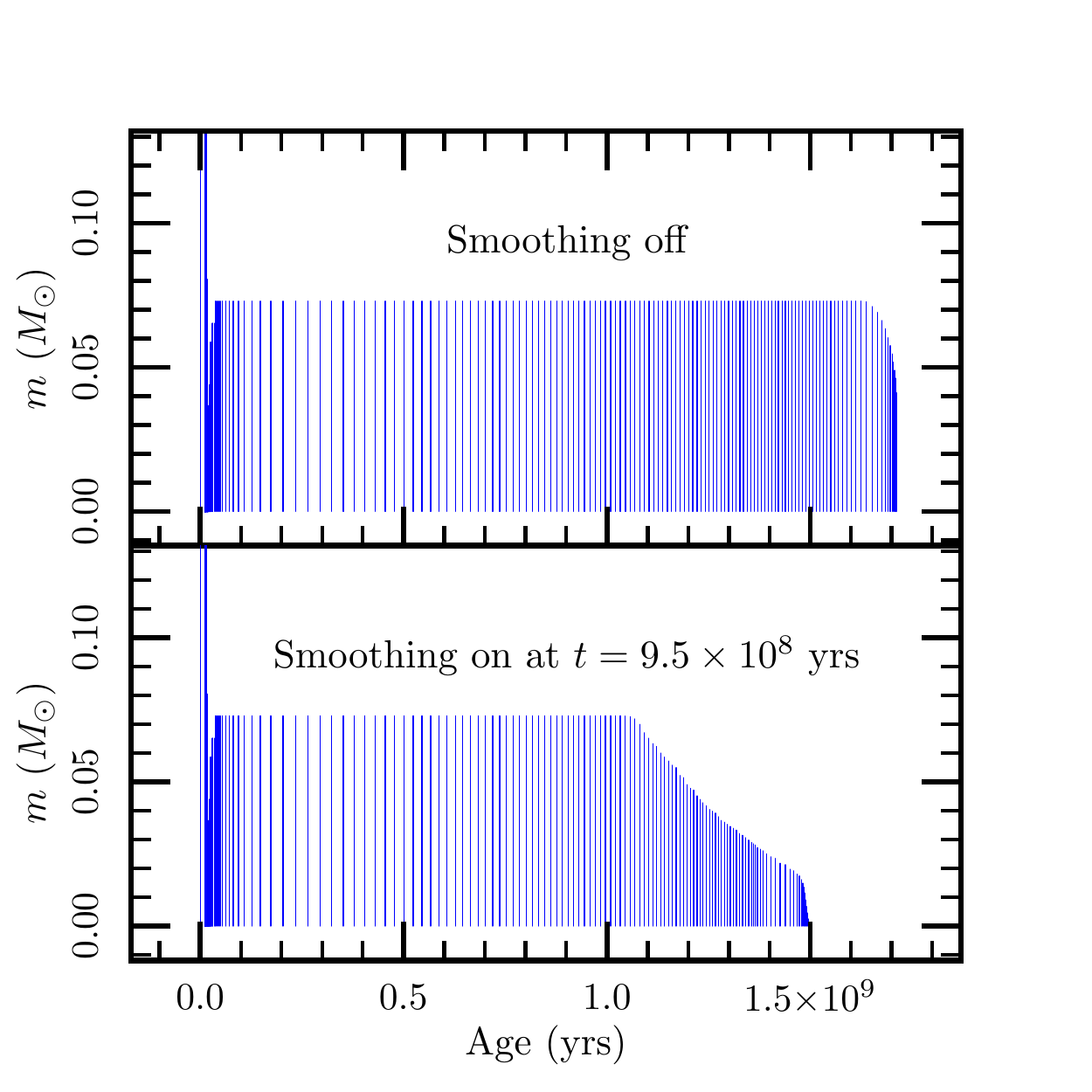}
	\caption{Comparison of the impact of $\nabla_\mu$-smoothing on convective core evolution of a $1.5\ M_\odot$ star. The top panel shows a Kippenhahn diagram of evolution under the Ledoux criterion without $\nabla_\mu$-smoothing, where the vertical lines mark the extent of convective regions at each time step. The bottom panel shows the evolution of the same star, but with $\nabla_\mu$-smoothing turned on suddenly (as in Figure \ref{fig:smoothtest}) when the core hydrogen mass fraction drops below $0.4$ which occurs at $t=9.5\times 10^8$ yrs. As soon as $\nabla_\mu$-smoothing is turned on, the core begins to shrink at a near constant number of cells per step. The apparent delay in core shrinkage in mass coordinate is merely due to the increased spatial resolution near the convective boundary.}
	\label{fig:kipp_smoothing}
\end{figure}

Figure \ref{fig:smoothtest} shows the evolution of a compositional discontinuity at the convective boundary in a numerical experiment in which compositional smoothing is suddenly switched on during the main sequence. This clearly illustrates how the spike in $\nabla_\mu$ that had built up at the convective boundary starts to diffuse inward when smoothing is turned on, causing the convective boundary to artificially move inward with it. Figure \ref{fig:kipp_smoothing} shows a Kippenhahn diagram of a similar run, as well as that of a star where no smoothing was used. As soon as smoothing is enabled, the convective core starts moving inward, resulting in a significantly smaller convective core for the remainder of the main sequence compared to the run without smoothing. This effect occurs for \textit{all} stars that develop a convective core - not just stars that show a strong difference between their evolution under the Ledoux and Schwarzschild criteria; smoothing $\nabla_\mu$ always shrinks a convective core relative to the same model without smoothing. We consider this to be an artificial effect and our models are run without $\nabla_\mu$-smoothing, unless otherwise noted. We will see in section \ref{sec:applications} that this distinction is rendered moot when layered semiconvection is accounted for because of the mixing that naturally occurs beyond the convective boundary.

\section{Applications to stellar evolution}
\label{sec:applications}

\subsection{Where can semiconvection make a significant impact on the evolution of a star?}
\label{subsec:where_semi}
As discussed in section \ref{sec:intro}, semiconvective mixing only occurs in the presence of both stable composition gradients and unstable thermal gradients, which are typically found in regions adjacent to convective zones. The most natural source of stabilizing composition gradients is nuclear burning, and
convective burning cores exist in main sequence stars (with $M\ge 1.2\ M_\odot$) as well as during core helium burning. 

\begin{figure}
	\plotone{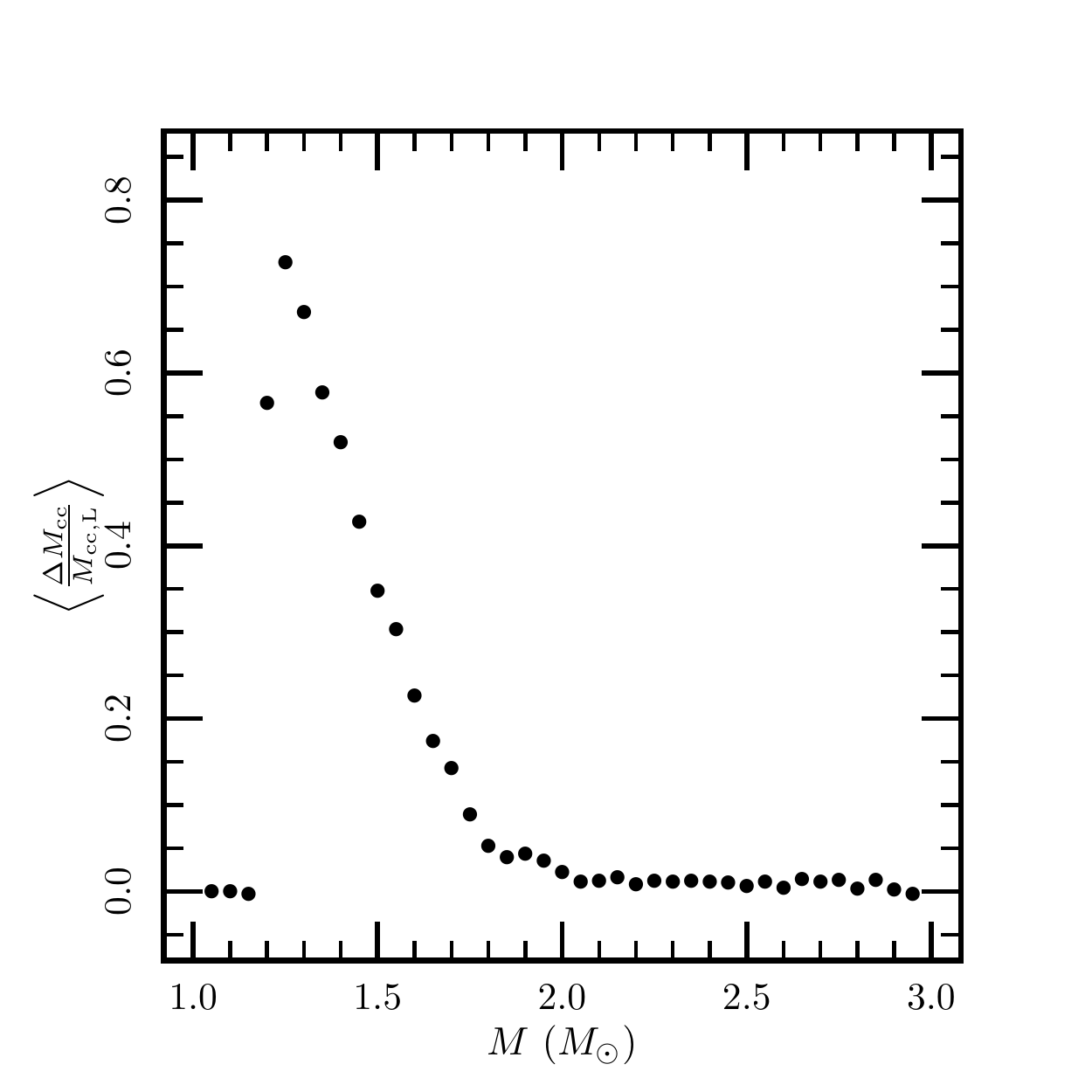}
	\caption{Measure of potential impact of semiconvection on the cores of main sequence stars of varying masses, as described in the main text. 
	Core convection occurs for stars with $M \ge 1.2\ M_\odot$, and the larger the computed ratio is, the larger the potential effect of semiconvection on the evolution of the star. A small window exists from $M \approx 1.2 - 1.7\ M_\odot$ where there is a large enough compositional gradient outside the convection zone for semiconvection to have a significant impact on the star. See Figure \ref{fig:grad_profile_comparison} for an illustration of the compositional gradients in selected stars.}
	\label{fig:mcc_vs_mass-low}
\end{figure}

In order to best understand the effects of our new semiconvective mixing prescription on stellar evolution, we first identify stars in which it may have a large effect. Figure \ref{fig:mcc_vs_mass-low} shows a measure of the potential impact of semiconvection on the evolution of main sequence stars in the $1-3\ M_\odot$ range. We quantify this impact by comparing the predicted convective core sizes computed without semiconvection (and without smoothing, see above) under the Schwarzschild and Ledoux criteria, respectively. Because the Schwarzschild criterion does not take into account compositional gradients while the Ledoux criterion does, we expect the sizes of convective cores calculated in the Schwarzschild case to be larger than in the Ledoux case in the presence of stabilizing compositional gradients. How much larger will depend on the size of $\nabla_\mu$ outside the convective core, which in turn depends on the local nuclear burning rates and varies with stellar mass. Any semiconvective region, should it exist, must necessarily reside in between the Schwarzschild and Ledoux core boundaries. Its maximum extent is therefore well approximated by the difference between the two radii, which can be used as a proxy for the potential impact of semiconvection.

Figure \ref{fig:mcc_vs_mass-low} shows the time-averaged ratio of the difference in convective core mass between the Schwarzschild and Ledoux cases, scaled to that of the Ledoux case, computed as
\begin{equation}
\left<\frac{\Delta M_{\rm cc}}{M_{\rm cc, L}} \right> = \frac{1}{\tau_{\rm ms} - \tau_0} \int_{\tau_0}^{\tau_{\rm ms}} \frac{M_{\rm cc, S} - M_{\rm cc, L}}{M_{\rm cc, L}}\ dt,
\end{equation}
where $M_{\rm cc, S}$ is the convective core mass obtained using the Schwarzschild prescription, $M_{\rm cc, L}$ is the convective core mass obtained using the Ledoux prescription, and the interval $[\tau_0, \tau_{\rm ms}]$ spans the period between the onset of core convection, and the end of the main sequence. Since this ratio is calculated from two different stellar models, the upper bound on the integral is taken to be the shortest lifetime - here the Ledoux model due to the smaller convective core. The advantage of this measure is that it does not depend at all on the semiconvective prescription employed, but merely probes which main sequence stars have significant chemical gradients outside of their convective cores.
Figure \ref{fig:mcc_vs_mass-low} shows that the most favorable mass range to explore the effects of semiconvection in main sequence stars is $M \approx 1.2 - 1.7\ M_\odot$. We have performed these calculations on stars with masses up to $30\ M_\odot$ without seeing another window where semiconvective mixing has a significant impact on convective core evolution during the main sequence.

\begin{figure}
	\plotone{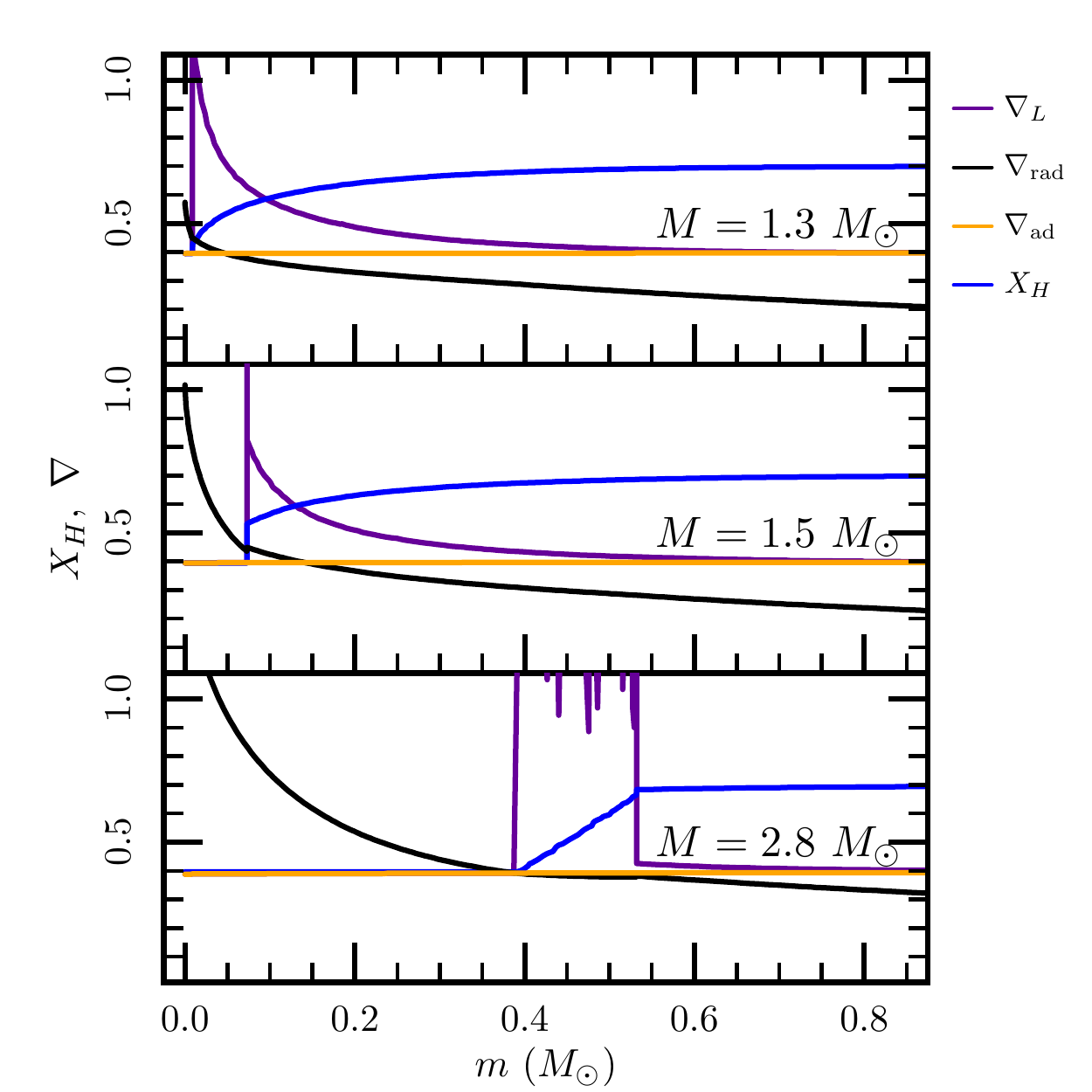}
	\caption{Profiles of composition gradients $\nabla_L$ (purple), radiative gradient $\nabla_{\rm rad}$ (black), adiabatic gradient $\nabla_{\rm ad}$ (orange), and mass fractions of hydrogen ($X_H$, blue), as a function of mass coordinate for stars that can have large semiconvective regions ($1.3\ M_\odot$ and $1.5\ M_\odot$) as well as a $2.8\ M_\odot$ star that cannot support significant semiconvection outside of its core. These profiles are obtained by evolving each star under the Ledoux criterion until the central mass fraction of hydrogen first drops below $0.4$. The source of the mass-dependence in where semiconvection can occur comes from the size of the chemical gradient term in $\nabla_L$ immediately outside the convective cores. Large values of $\nabla_L$ outside small convective cores in stars in the mass range $1.2 - 1.7 \ M_\odot$ cause a significant difference between the convective boundaries determined by the Ledoux and Schwarzschild criteria: the former is the position of the $\nabla_L$ spike, located at $m/M_\odot = 0.01, 0.07,$ and $0.38$, respectively. The latter is the position where $\nabla_{\rm rad} = \nabla_{\rm ad}$, located at $m/M_\odot = 0.04, 0.14,$ and $0.38$, respectively. For higher mass stars such as the $2.8\ M_\odot$ case shown, the compositional gradient outside the core is too small to allow for large semiconvective regions.}
	\label{fig:grad_profile_comparison}
\end{figure}

The significance of this particular mass range for semiconvection is best understood by looking at the compositional gradient generated by nuclear burning outside the convective core. Figure \ref{fig:grad_profile_comparison} shows the difference between stars that are susceptible to semiconvection ($1.3\ M_\odot$ and $1.5\ M_\odot$ models), and a $2.8\ M_\odot$ model which is not. The relatively weak temperature dependence of the proton-proton chain allows burning to occur in regions outside of the convective core at a rate which gradually drops with radius due to the temperature and density decreasing outwards. This radius-dependent burning rate causes the development of a composition gradient which steepens inward. While all stars in the mass range shown in Figure \ref{fig:mcc_vs_mass-low} have nearly the same $\mu$-profile outside their core for a given age, the position of the edge of the core is strongly dependent on the stellar mass and is located at radii with stronger or weaker $\mu$-gradients. Lower-mass stars with smaller convective cores have larger $\mu$-gradients just outside their cores, while larger-mass stars with larger cores have correspondingly weaker $\mu$-gradients. This is why semiconvection can make the largest difference in the evolution with stars having the smallest convective cores.


Stars may also be susceptible to semiconvection during core helium burning phases. However, helium burning reactions are much more temperature sensitive and do not extend far out of the convective core. As a result, there is not a large difference between the sizes of convective cores calculated under different convective criteria. A compositional gradient can also exist outside of the hydrogen burning shell due to the less temperature-sensitive proton-proton chain burning outside of the main CNO burning region, but this effect is much less dramatic than during the main sequence due to the higher temperatures in the burning regions, and to the much shorter remaining stellar lifetime. We defer a discussion of the impact of semiconvective mixing on the later stages of high-mass stars to a future paper.

\subsection{Maintaining a composition gradient}
\label{subsec:maintaining_gradient}
In order for semiconvective regions to persist for significant periods of time, their turbulent mixing rates cannot be too large, otherwise the compositional gradient that causes them to exist will be destroyed. If this happens, the semiconvective region is simply converted into a convective region since it remains Schwarzschild-unstable. The efficiency of compositional mixing in semiconvective regions is determined by the size of the diffusion coefficient $D_{\rm semi}$ (see equation \ref{eq:d_layered}) and therefore by the layer height $H_L$ - larger layer heights imply larger diffusion coefficients. This suggests that stellar models with layered semiconvection can be split into three groups - (a) one in which the compositional mixing is too weak to have any influence on stellar evolution, (b) one in which the convective core size is modified, but the mixing is slow enough for the semiconvective regions to persist throughout the main sequence, and (c) one where the compositional mixing is fast enough to turn semiconvective zones into convective zones over a time scale much shorter than the main sequence lifetime. We anticipate the existence of a critical layer height, $H_{\rm L, crit}$, such that evolution with $H_{\rm L} \gg H_{\rm L, crit}$ (case c) is similar to one without semiconvection, but where the convective criterion is given by the Schwarzschild criterion. Meanwhile, if $H_{\rm L} \ll H_{\rm L, crit}$ (case a) we expect the star to evolve as if semiconvection was absent, but the convection is determined by the Ledoux criterion. 

\begin{figure}
	\plotone{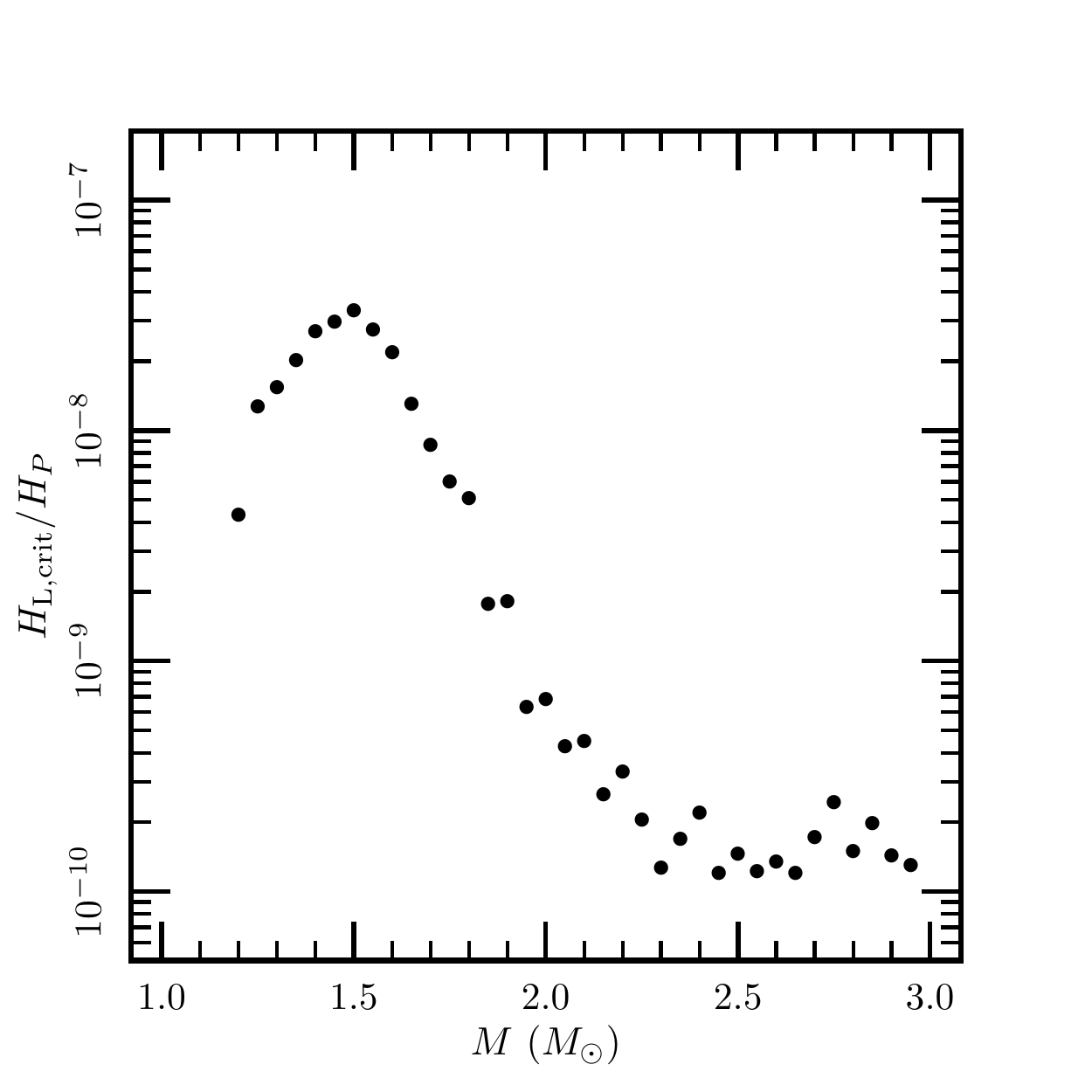}
	\caption{Estimates of the time-averaged critical layer height, $H_{\rm L, crit}$, during main sequence evolution as a fraction of pressure scale height as a function of stellar mass. Semiconvective layer heights larger than $H_{\rm L, crit}$ will rapidly mix the semiconvective regions before the star evolves off the main sequence, destroying the chemical gradient that supports them and turning the region convective. Smaller layer heights will allow layered semiconvective regions to maintain their compositional gradient and survive through the main sequence. For all masses considered here, the critical layer height is orders of magnitude smaller than the expected minimum layer height, which suggests that layered semiconvection cannot persist in main sequence stars and convective core evolution is therefore well approximated with that obtained under the Schwarzschild criterion.}
	\label{fig:hlayer_crit}
\end{figure}

When semiconvection is enabled and material can be exchanged between the convective core and surrounding regions, there is a competition between nuclear burning which maintains the compositional gradient and semiconvective mixing which tries to remove the gradient. We can therefore estimate $H_{\rm L, crit}$ by comparing the mixing timescale of the semiconvective region to the compositional evolution time scale in the convective core. The semiconvective mixing timescale is given by
\begin{equation}
t_{\rm semi} = \frac{l_{\rm semi}^2}{\left<D_{\rm semi}\right>},
\end{equation}
where $l_{\rm semi}$ is the radial extent of the entire semiconvective region and $\left<D_{\rm semi}\right>$ is the mass-averaged diffusion coefficient in the semiconvective region. Similarly, we can compute the compositional change timescale as
\begin{equation}
t_{\rm \mu, center} = \left| \frac{\mu_{\rm center}}{\dot{\mu}_{\rm center}} \right|,
\end{equation}
where $\mu_{\rm center}$ is the mean molecular weight of the material at the center of the star. This is the same as the $\mu$-evolution time scale for the entire convective core since it is fully mixed. Changing $H_L$ affects the value of $t_{\rm semi}$, but not $t_{\rm \mu, center}$. The critical layer height is the value of $H_L$ for which these two time scales are equal. Writing the diffusion coefficient as
\begin{equation}
D_{\rm semi} = D_{\rm semi, 0}\ \left(\frac{H_{\rm L}}{H_P}\right)^{1.48} \approx D_{\rm semi, 0}\ \left(\frac{H_{\rm L}}{H_P}\right)^{3/2},
\end{equation}
we can then estimate $H_{\rm L, crit}$ by equating $t_{\rm semi}$ and $t_{\rm \mu, center}$,
\begin{equation}
H_{\rm L, crit} \approx H_P \left(\frac{l_{\rm semi}^2}{D_{\rm semi,0}\ t_{\rm \mu,center}}\right)^{2/3}.
\end{equation}
In order to estimate $l_{\rm semi}$, we run a model under the Ledoux criterion and calculate where the boundary of the convective core would be under the Schwarzschild criterion; $l_{\rm semi}$ is the distance between the actual Ledoux and hypothetical Schwarzschild convective boundaries \footnote{This is the same as the size of semiconvective regions in models where chemical transport is turned off (eg. by setting $D_{\rm semi} = 0$) or the layer heights used are small enough where mixing is inconsequential.}.
Figure \ref{fig:hlayer_crit} shows the resulting time-averaged $H_{\rm L, crit}$ values over the main sequence evolution of stars in the mass range of $1.2-3.0\ M_\odot$. Given that $H_P \approx 10^{10}$ cm near the cores of such stars, we find that $H_{\rm L, crit}$ ranges from $\approx 10^2 - 10^3$ cm for stars in the mass range of greatest potential semiconvective impact, $1.2-1.7\ M_\odot$.

While $H_{\rm L, crit}$ was determined from basic timescale arguments, we can easily compare it to the actual layer heights realized in numerical simulations. \citet{Wood13} found that the minimum layer height is about $10$ times the scale of the basic instability, $l_{\rm ODDC}$, with
\begin{align}
l_{\rm ODDC} = 10d &= 10\left(\frac{\kappa_T \nu}{g \delta \left| \der{\ln T}{r} - \der{\ln T^{\rm ad}}{r} \right|}\right)^{1/4} \nonumber \\
&= 10 \left(\frac{P \kappa_T \nu}{\rho g^2 \delta \left| \nabla - \nabla_{\rm ad} \right|}\right)^{1/4} \approx 10^4\ {\rm cm}.
\end{align}
This gives a minimum layer height of $\approx 10^5$ cm, much larger than $H_{\rm L, crit}$. 
This has a fundamental consequence: layered semiconvection is so efficient in main sequence stars that is accurately approximated by evolution under the Schwarzschild criterion, ignoring semiconvection altogether!

\begin{figure}
	\plotone{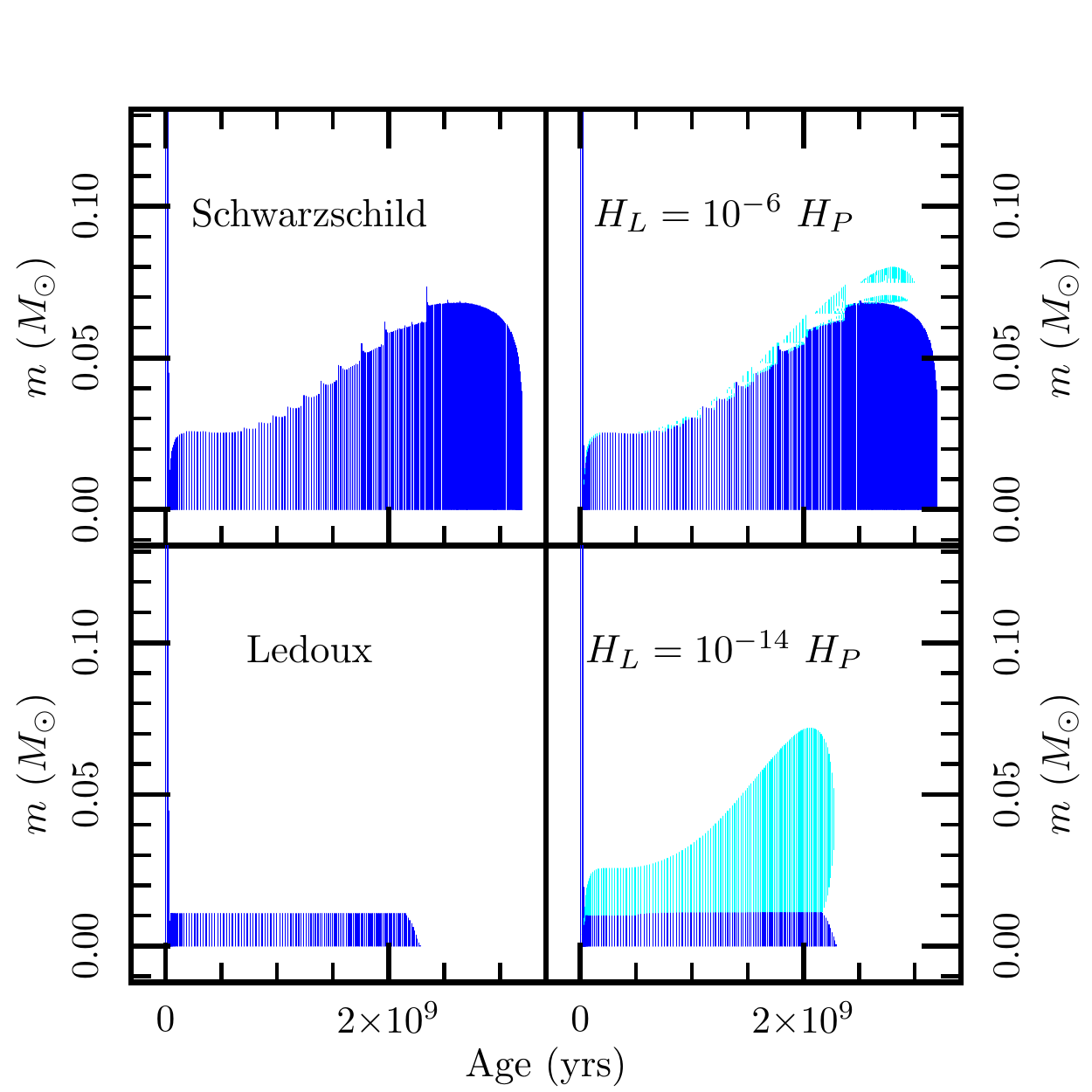}
	\caption{Kippenhahn diagrams for the main sequence evolution of a $1.3\ M_\odot$ star using different mixing criteria. The panels on the left show the evolution of the convective core (blue) under the Schwarzschild and Ledoux criteria. The convective core is much smaller in the Ledoux case because there is a significant stabilizing compositional gradient outside the core due to pp-chain burning. The main-sequence lifetime of this case is therefore significantly shorter. The panels on the right show the corresponding evolution with layered semiconvection using layer heights many orders of magnitude larger and smaller than the critical layer height, $H_{\rm L, crit} \approx 10^{-8}\ H_P$. The core evolution of the larger $H_L$ case is virtually the same as one obtained with a model using the Schwarzschild criterion, while the evolution under the smaller $H_L$ case is virtually the same as the Ledoux case. Only the $H_L \ll H_{\rm L, crit}$ model shows large semiconvective zones (light blue) over the entirety of the main sequence, because the mixing is too slow to remove the compositional gradient.}
	\label{fig:1_3msun_kipp}
\end{figure}

\begin{figure}
	\plotone{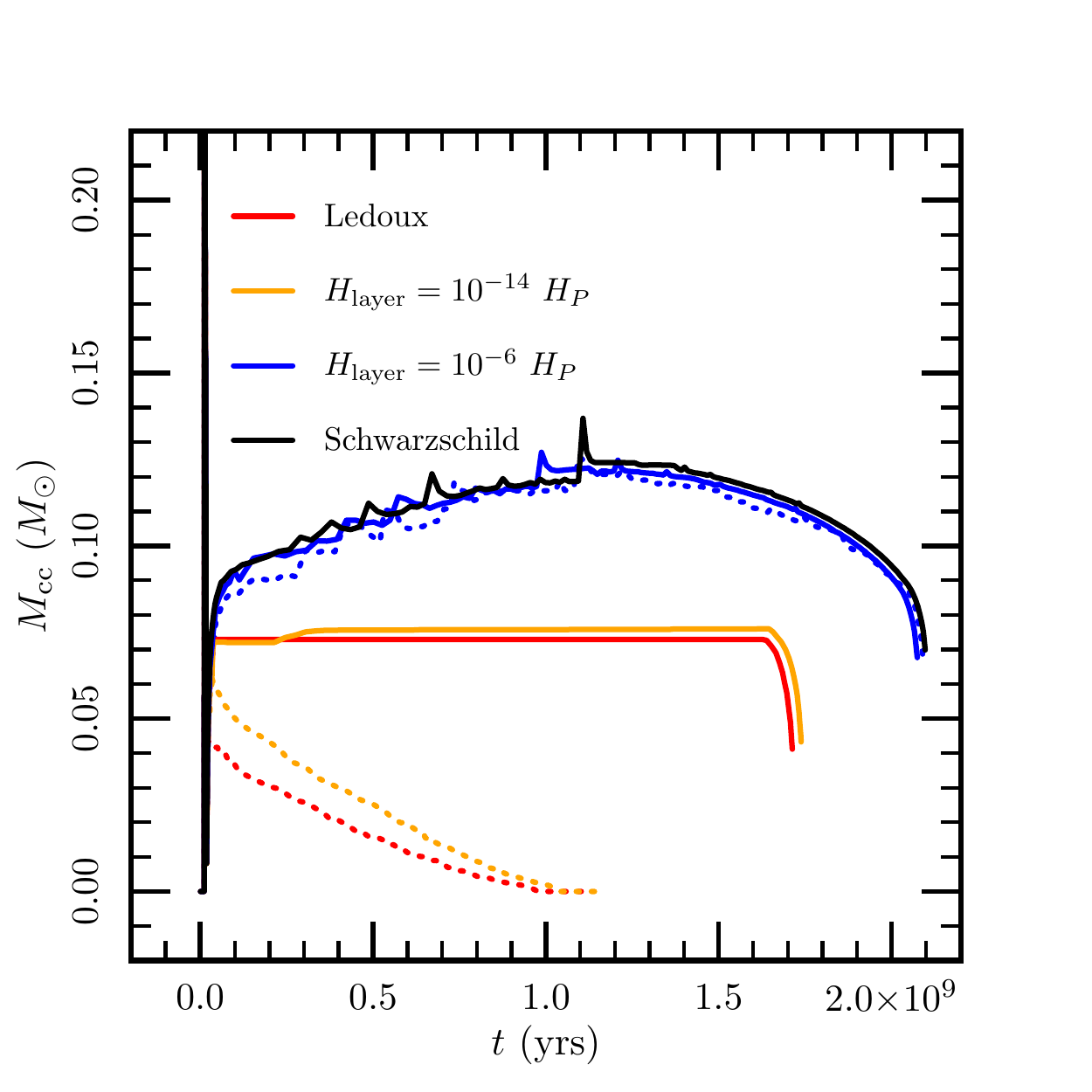}
	\caption{Main sequence evolution of the convective core mass, $M_{\rm cc}$, for $1.5\ M_\odot$ stars under various mixing schemes. Dashed lines indicate the same mixing parameters as solid lines, but with the default $7$-point Gaussian $\nabla_\mu$-smoothing enabled. By adjusting the convection criterion, smoothing, and strength of layered semiconvection, the convective core evolution falls into roughly three cases. The smallest convective cores occur when the Ledoux criterion (or layered semiconvection with with $H_L \ll H_{\rm L, crit}$) is used in conjunction with $\nabla_\mu$-smoothing. Mid-size convective cores occur for the same cases but without $\nabla_\mu$-smoothing enabled. The largest convective cores occur with either the Schwarzschild criterion or with layered semiconvection where $H_L \gg H_{\rm L, crit}$. For those cases, whether $\nabla_\mu$-smoothing is enabled or not does not have a significant effect since the mixing is strong enough to destroy compositional gradients.}
	\label{fig:Mcc_1_5msun}
\end{figure}

Figure \ref{fig:1_3msun_kipp} illustrates this statement with Kippenhahn diagrams of a $1.3\ M_\odot$ star evolved with various mixing prescriptions. Although unphysically small, we can investigate the effect of layered semiconvection using layer heights $H_{\rm L} \ll H_{\rm L, crit}$. As expected, we find that semiconvective regions are effectively radiative and the star evolves as if the Ledoux criterion was used. For the more realistic case where $H_{\rm L} \gg H_{\rm L, crit}$, the evolution is nearly identical to the Schwarzschild case, as predicted above. Our results also highlight how sensitive the evolution is to the layer height. There are several orders of magnitude difference in $H_L$ between models that are effectively the same as the Schwarzschild case and models that are effectively the same as the Ledoux case. Intermediate values of $H_L$ initially have semiconvection zones of the same size that eventually transition into convective zones before core hydrogen depletion.
 Figure \ref{fig:Mcc_1_5msun} summarizes the convective core evolution for several different mixing criteria, showing convective cores that fall into three size groups. The smallest convective cores occur when smoothing is enabled, and semiconvective mixing is either turned off or is weak enough ($H_L \ll H_{\rm L, crit}$) that the evolution is effectively Ledoux. Convective cores are larger for the same cases when smoothing is not used. Finally, the largest convective cores occur for efficient semiconvection (the only physically realizable outcome, where $H_L \gg H_{\rm L, crit}$) and is nearly identical to the Schwarzschild case. For these models, turning on smoothing does not have a noticeable effect due to the strong mixing outside the convective boundary.

\begin{figure}
	\plotone{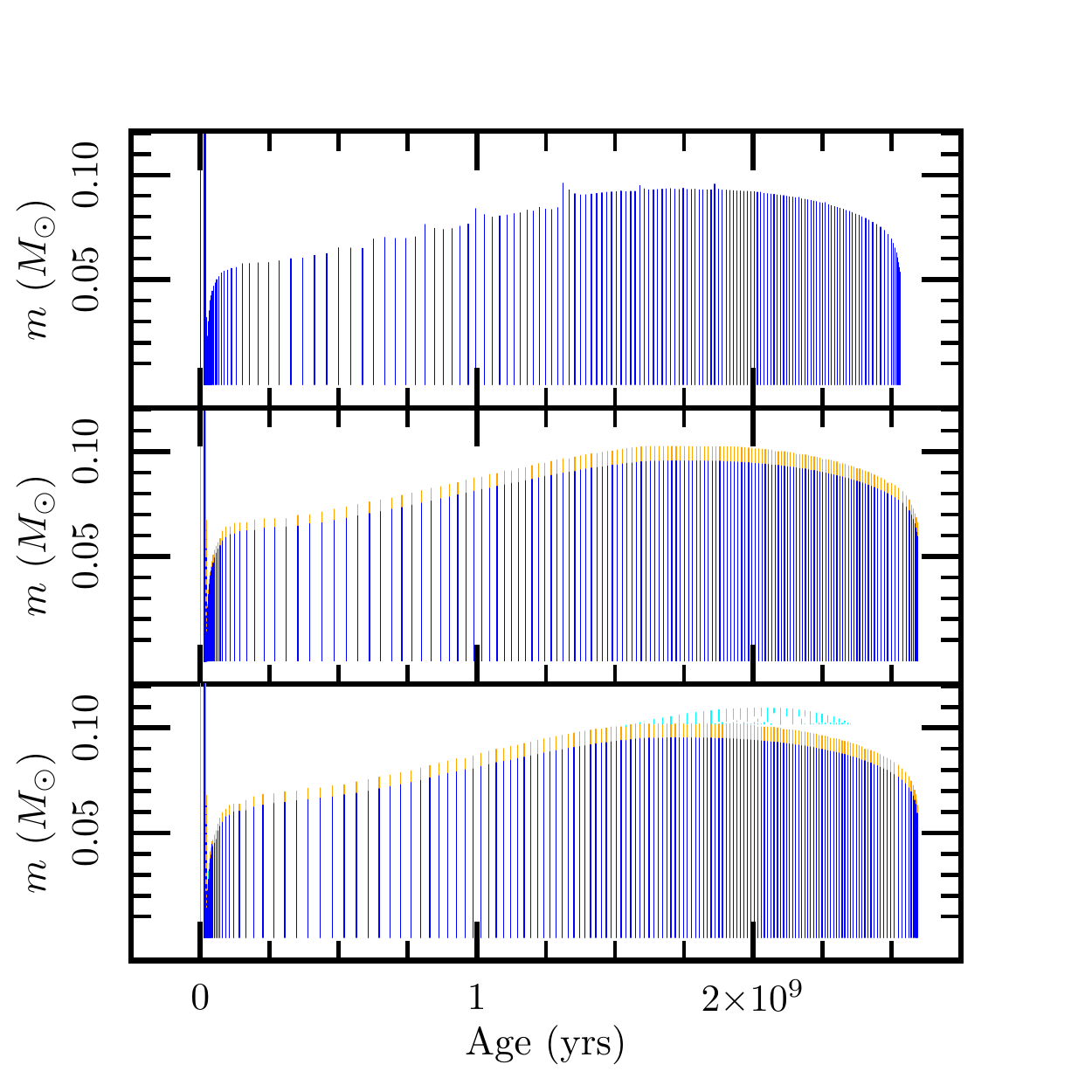}
	\caption{Kippenhahn diagrams for the main sequence evolution of a $1.4\ M_\odot$ star showing the effects of overshoot on smoothing out the temporal evolution of the convective core. Dark blue regions are convective, light blue are semiconvective, and orange correspond to overshooting. The top panel shows the evolution under the Schwarzschild criterion. The middle panel shows the same evolution but also with overshoot turned on (using $f=f_0=10^{-4}$). The bottom panel shows evolution with layered semiconvection at a value of $H_L = 10^{-6}\ H_P$ and the same overshooting parameters.}
	\label{fig:kipp_rows_1_4msun_overshoot.pdf}
\end{figure}

Finally, we show that the same results hold in models with convective overshoot. The spikiness in the convective core mass under Schwarzschild evolution can be removed with a small amount of convective overshoot. Figure \ref{fig:kipp_rows_1_4msun_overshoot.pdf} shows the evolution of a pure Schwarzschild model, a Schwarzschild model with overshoot, and a semiconvective model with $H_L \gg H_{\rm L, crit}$ (thus effectively as if they were run with the Schwarzschild criterion). Semiconvective models that are effectively Schwarzschild models remain so with convective overshoot, so the respective prescriptions combine without issue.


\section{Conclusion}
\label{sec:conclusions}

Convective core evolution of main sequence stars in the mass range $1.2 - 1.7\ M_\odot$ can be dramatically impacted by semiconvective mixing due the extended compositional gradient outside the convective core coming from nuclear burning. We investigated the effects of the layered semiconvection prescription given in \citet{Wood13} on the evolution of such stars, finding that there is a critical layer height, $H_{\rm L, crit}$, above which the evolution is effectively given by ignoring compositional gradients altogether and evolving the star using the Schwarzschild criterion. For layer heights smaller than $H_{\rm L, crit}$, the evolution is effectively given by ignoring the additional mixing beyond the convective core and evolving the star using the Ledoux criterion. This critical layer height is orders of magnitude smaller than the minimum layer height predicted from the underlying instability, so if layered semiconvection occurs within stars, we expect it to be very effective at mixing composition and quickly erasing the compositional gradient that allows it to exist. Such a star evolves as if the Schwarzschild mixing criterion were employed. We also found that numerically smoothing the compositional gradient term can significantly change the sizes of convective cores. When using the Ledoux criterion, this $\nabla_\mu$-smoothing will artificially push the convective core boundary inwards, shrinking the convective core. This effect can be dramatic if no additional mixing is included beyond the convection zone, and can artificially reduce the lifetimes of evolutionary phases with convective cores by up to fifty percent.

It may be possible in principle to infer the sizes of convective cores in main sequence stars from observations of pulsation modes \citep{Mazumdar06, Cunha07, Silva-Aguirre10b, Brandao14}. However, detecting solar-like oscillations in other main sequence stars is difficult and the current KEPLER and CoRoT data for such stars is much more limited than for brighter objects such as red giants. Sample sizes of solar-like oscillators (both main sequence stars and sub giants) are only in the dozens \citep{Appourchaux12, Metcalfe14}, while those of red giants are in the tens of thousands \citep{Stello13}. The few inferences of convective core sizes in main sequence stars show evidence of mixing beyond the Schwarzschild convection boundary, typically interpreted as a constraint on the amount of overshooting \citep{Silva-Aguirre13, Liu14, Deheuvels15}. It therefore may not be possible to place constraints on the strength of semiconvection in main sequence stars, so investigating its effect on evolved stars may be more fruitful. We are planning a future paper to examine this prospect.

We thank Justin Brown, Chris Mankovich, and Bill Paxton for useful discussions. Code and inlists necessary to reproduce our models are hosted on mesastar.org. This work was supported under grants NSF AST 0847477 and NSF AST 1211394.


\newpage
\bibliographystyle{apj}
\bibliography{Semiconvection}

\end{document}